\documentclass[journal]{IEEEtran}
\IEEEoverridecommandlockouts
\usepackage[pdftex]{hyperref}
\usepackage{amsfonts}
\usepackage{amssymb}
\usepackage{graphicx}
\usepackage{color}
\usepackage{amsthm}
\usepackage{amsmath}

\newcommand{\g}[1]{\ensuremath{\mathcal{#1}}}

\begin{document}

\title{Spiking label propagation for community detection}
\author{Kathleen~E.~Hamilton~\IEEEmembership{}
        and~Travis~S.~Humble~\IEEEmembership{}
\thanks{K. E. Hamilton and T. S. Humble are with the Quantum Computing Institute, Oak Ridge National Laboratory, Oak Ridge, Tennessee 37831-6015 USA e-mail: hamiltonke@ornl.gov.}
\thanks{Manuscript received XXX; revised YYY.}
\thanks{This manuscript has been authored by UT-Battelle, LLC, under Contract No. DE-AC0500OR22725 with the U.S. Department of Energy. The United States Government retains and the publisher, by accepting the article for publication, acknowledges that the United States Government retains a non-exclusive, paid-up, irrevocable, world-wide license to publish or reproduce the published form of this manuscript, or allow others to do so, for the United States Government purposes. The Department of Energy will provide public access to these results of federally sponsored research in accordance with the DOE Public Access Plan.}
}
\maketitle

\begin{abstract}
In this paper we present spike-based label propagation for community detection in undirected graphs. Using a fully connected system of deterministic spiking neurons driven by external currents, the generated spike responses are decoded into binary signals and the Hamming distance between pairs of signals is used to group vertices into communities.  We test our approach on a set of graph instances of $128$ vertices and either homogeneous or heterogeneous community size distributions. Our results are presented as proof-of-concept that spiking neurons can be incorporated into graphical analysis tasks, and as a demonstration of a  computing workflow that can utilize neuromorphic hardware without extensive pre-training of network parameters.  
\end{abstract}

\maketitle

\section{Introduction}
Computing with pulsed (or spiking) neural networks \cite{maass1997networks,maass2001pulsed} has advanced in recent years with the development of neuromorphic hardware \cite{indiveri2015memory,merolla2014million,painkras2013spinnaker,davies2018loihi}.  Two main applications have been: biological modeling, and the adaptation of deep learning algorithms to incorporate spiking signals. The derivation of spiking-based back propagation methods \cite{esser2015backpropagation, stromatias2015scalable}, and spike-based Gibbs sampling methods \cite{das2015gibbs} has resulted in the successful adaptation of convolutional neural networks \cite{esser2016convolutional}, restricted Boltzmann machines \cite{pedroni2016mapping} and recurrent neural networks \cite{diehl2016conversion} for neuromorphic systems.  However, these applications require significant computational overhead to pre-train neural networks before they are then adapted into a spike-based formalism.

There is growing interest in applications for neuromorphic hardware that fall outside the realm of deep learning or biological modeling.  Spike-based annealing methods have been developed in order to solve optimization problems, such as the Traveling Salesman, graph coloring and MAX-SAT \cite{constraintSNN_Maass_2016,constraintSNN_SpiNNaker_2017} on neuromorphic hardware.   Additionally, there have been recent studies showing spiking neurons can be used for Markovian random walks \cite{severa2018spiking} and general scientific computing \cite{severa2016spiking}. 

Community detection is a difficult problem to solve, but its ubiquity in many scientific disciplines has lead to the proliferation of methods \cite{boccaletti2006complex,fortunato2010community,malliaros2013clustering,schaub2017many}
 and algorithms \cite{xu2005survey}.  Correlations play an important role in label propagation \cite{raghavan2007near,barber2009detecting} and other Potts model-based methods \cite{tibely2008equivalence,PhysRevLett.76.3251,PhysRevE.74.016110,reichardt2004detecting}.  An advantage of methodos based off of interacting spin models, is the ability to identify groups of related vertices in a graph without extensive prior knowledge of the underlying community distribution.  
 
A spiking neuron will fire along all of its outgoing synapses and we believe the incorporation of this isotropic behavior can potentially speed up graphical analysis.  In this work we define spike-based label propagation (SLP) which incorporates correlated dynamics of spiking neurons into community detection for undirected graphs.   Previous studies of SLP have focused on how the collective dynamics of spiking neuron populations can lead to large scale synchronization \cite{quiles2010label,quiles2013dynamical} using systems of leaky-integrate and fire neurons and also Kumamoto oscillators \cite{de2014community}.  Our approach uses leaky integrate and fire neurons and propagates labels through the network based on local correlations between spiking neurons.  However, our approach relies on the identification of specific spike dynamics caused by applying external stimulus to a single neuron.  We do not use a global inhibitor function, instead a high density of inhibitory synapses control and limit the global spike dynamics.  

\section{Definitions}
\label{sec:mapping}
We identify communities in unweighted graphs with no self-loops or multiple edges. A graph $\g{G} = \g{G}(V,E)$ is fully defined using a vertex set $V(\g{G})=\lbrace v_i\rbrace$ and a set of symmetric connections ($E(\g{G}) = \lbrace e_{ij}\rbrace$, $e_{ij} = e_{ji} = (v_i, v_j)$). A spiking neuron system (SNS) is similarly defined $\g{S} = \g{S}(N_{k},W_{\ell})$; using a neuron set ($N_{k} = \lbrace n_i \rbrace $, $k = (v_{th},t_{R},\tau)$), and a set of synaptic connections ($W_{\ell}=\lbrace w_{ij} \rbrace $, $w_{ij} = w_{ji}= (n_i,n_j)$, $\ell = (s_w) $).  The choice of parameters is dependent on the underlying spiking neuron model and is further explained in Sec. \ref{sec:LIF_model}. 

Each neuron has a time dependent internal state which is quantified by a membrane potential ($n_i(t)$). This potential function controls the spiking behavior of the neuron: if $n_i(t) < v_{th}$ then no spike will be detected,  if $n_i(t) \geq v_{th}$ a spike is fired. The synaptic weights control the effect an arriving spike has on the membrane potential: the arrival of a spike causes a change in potential $\Delta n = s_w$. The output of a spiking neural system is a \textit{spike raster}, which is a recording of when every neuron fires a spike. For each neuron a \textit{spike train} is defined by all the times that neuron fires a spike over the course of a simulation.  

A spin glass mapping takes a graph $\g{G}(V,E)$ to a fully connected SNS $\g{S}(N,W)$: every vertex $v_i$ to a neuron $n_i$ ($|V| = |N|$) and every edge in $E(\g{G})$ to a positive-weighted synapse $e_{ij} \to w_{ij}\in W(\g{S}), s_w>0$. Any edge absent in \g{G} is created in $W(\g{S})$ as an negative-weighted synapse $e_{ij} \notin E(\g{G}) \to w_{ij} \in W(\g{S}), s_w < 0$.  Spikes are transmitted between adjacent neurons via the weighted synaptic connections, and we assume that the transmission is done without attenuation.  As a result: strongly connected vertices of a graph will map to groups of neurons which are strongly connected by positively weighted synapses.   Neurons in the same community are identified by a high degree of similarity in their binary spike vectors; a converse of the well-known aphorism in Hebbian learning \cite{Lowel209},``if neurons fire together, they should wire together.''  

\subsection{Spiking neuron model}
\label{sec:LIF_model}
We choose to build our method with leaky-integrate and fire (LIF) neurons and ideal synaptic connections due to their use in near-term neuromorphic hardware \cite{merolla2014million,cassidy2013cognitive}.  While there are many parameters that are incorporated into this model, we define our neurons using as few parameters as possible.  A neuron in our model is fully defined by its spiking threshold $v_{th}$, rest voltage $v_0$ and decay rate $\tau$. The membrane potential of each neuron, is described by the differential equation:
\begin{align}
\dot{n}_j&= \frac{\left(J_{ext}(t) - n_j (t)\right)}{\tau},\\
J_{ext}(t) &= I_{ext}(t)R + \sum_{t^{(f)}}\sum_{\substack{i \to j,\\ i \neq j}} w_{ij} \delta(t - t^{(f)}).
\label{eq:LIF_EOM}
\end{align}
The state of a neuron can change in time due to the decay of any accumulated charge, by an external driving current $I_{ext}$ or the arrival of a spike.  When a spike arrives at neuron $n_i$ from neuron $n_j$ at time $t=t_f$ alters the state $n_i(t_f)$ by $w_{ij}$; where $\delta(t-t_f)=1$ at the firing time of a neuron, and the synaptic weight $s_w$ associated with the synapse $w_{ij}$ can be positive or negative. We include a unit resistance factor $R=1 \: \mathrm{\Omega}$ to ensure the first term of $J_{ext}(t)$ has the appropriate units. 

\subsection{Generation of spike trains}
\label{sec:spike_driving}
We generate spiking dynamics by applying external currents to the SNS.  At any time during the system evolution, only one neuron is driven by an external current, but depending on the neuron parameters (firing threshold, time constant) and synapse weight $s_w$, it is possible that correlated spiking responses can be generated by spikes traveling between neurons strongly connected by positive synapses while simultaneously suppressing spiking responses between weakly connected neurons. 

When an external driving current is applied, Eq. \eqref{eq:LIF_EOM}, $J_{ext}(t)$ contains the external current term,
\begin{equation}
\dot{n}_j = \frac{1}{\tau}\left(I_{ext}(t)R - n_j (t) \right).
\label{eq:primary_dynamics}
\end{equation}
We use external currents that generate uniform spike patterns (spike emitted at a uniform firing rate).  The simplest external current that leads to uniform spike firing is a constant external current $I_{ext}(t) = I$. In order to generate uniform spiking in a discrete time interval the spin glass spiking neural networks are driven by a sequence of Heaviside step currents (approximated by hyperbolic tangent functions). Each neuron in the system is driven separately by an external current term: 
\begin{equation}
I_{ext}(t)R = \dfrac{A_{max}}{2} [\tanh{(\beta [t-t_1])} - \tanh{(\beta[t_2-t])}].
\end{equation}
The pulse height $A_{max}$ is chosen such that external current is strong enough to rapidly drive a neuron across its firing threshold and the constant $\beta$ is chosen such that when the pulse is turned on the external potential quickly reaches $A_{max}$. Ensuring a minimal rise time for each square pulse also reduces the overlap between subsequent pulses applied to different neurons, and any potential change caused by $I_{ext}$ is assumed to fully decay from a driven neuron before any other neuron is driven. Nonetheless, when simulating the dynamics of spiking neurons, the use of numerical integration introduces the possibility that a spike can be fired before $I_{ext}R = A_{max}$.  Finally, the pulse width $t_A = t_2-t_1$ is chosen such that the driven neuron will fire multiple spikes to its neighbors and can generate multiple secondary spikes. 

When a neuron $n_i$ is being driven by an external current, it is sending weighted spikes to its neighbors. In Ref. \cite{Hamilton_NCAMA}, we discussed how to balance the driving parameters, and the neuron model parameters such that the arrival of $2$ positively weighted spikes at neuron $n_j$ will lead to the firing of a spike. In Eq. \eqref{eq:LIF_EOM}, $I(t)$ contains only the synapse term,
\begin{equation}
\dot{n}_j =  \frac{1}{\tau} \left(\sum_{t^{(f)}}\sum_{\substack{i \to j,\\ i \neq j}} w_{ij} \delta(t - t^{(f)})- n_j (t) \right).
\end{equation}
Once a primary spiking neuron has fired a neuron, it will enter a refractory period, during which of length $t_R$; any arriving spikes or external current will not have any effect on the membrane potential.  Under square pulse driving, the dynamics of a spin glass SNS is decoupled, neurons fire either due to an external current, or due to the arrival of positively weighted spikes, but never both. 

\subsection{Spike train decoding}
\label{sec:binary_decoding}
Binary decoding of a spike train converts the spike times over a duration of time $T$ into a binary vector $x_i$. From the spike raster, we convert each spike train to a binary vector $x_i^{m}$ of length $L = T/\Delta t$, by first fixing a time window $\Delta t$. The $k^{th}$ vector entry $x_i(k)=1$ if at least one spike occurs during this time window $[t_k, t_k+ \Delta t]$, otherwise the vector entry is $0$. The resulting binary vector $x_i^{(m)}$ can also include the (known) community label $(m)$.  Using the binary decoded spike trains, the degree of similarity between pairs of neurons is quantified using a Hamming distance metric \cite{humphries2011spike}.  This metric is,
\begin{equation}
H(x_i, x_j; \Delta t) = (1 - \frac{h(x_i,x_j)}{L}),
\label{eq:Hamming_metric}
\end{equation}
where $h(x_i,x_j)$ is the Hamming distance between two $n$-length binary vectors $x_i,x_j$.  If $H(x_i,x_j)\to 1$, then two spike trains are considered ``similar,'' but if $H(x_i,x_j)\to 0$ then two spike trains are considered ``dissimilar.'' In the case where either $x_i$ or $x_j$ is an inactive (non-firing) neuron we hard-wire in the condition that $H(x_i,x_j)=0$.  For the remainder of this work we will refer to Eq. \eqref{eq:Hamming_metric} as the Hamming similarity and use it to discern which spiking neurons (and vertices) are in the same community.

Though the width of the time window ($\Delta t$) does not appear explicitly in Eq. \eqref{eq:Hamming_metric}, it parameterizes the Hamming similarity by determining the length and weight of each binary vector $x_i$. If $\Delta t$ is narrow, such that each window can only contain a single spike, then the weight of the binary vector $\| x_i \|_1$ is maximized, however the overall vector length will grow and as a result the factor $h(x_i,x_j)/L$ can become very small. Similarly, if $\Delta t$ is wide and each window contains multiple spikes, the vector length $n$ can become very short, but the Hamming distance between vectors decreases. Again, as a result the factor $h(x_i,x_j)/L$ can become very small. In Sec. \ref{sec:discussion} we further discuss how it is necessary to balance maximizing $h(x_i,x_j)$ while minimizing $L$.

\section{Spiking label propagation}
\label{sec:spiking_label_propagation}
The similarity measure in Eq. \ref{eq:Hamming_metric} is used to incorporate the binary decoded spike trains $\lbrace x_i\rbrace$ into SLP. In Ref. \cite{raghavan2007near}, labels are propagated according to ``consensus'' between neighboring vertices. For binary decoded spike trains, we assign the same community labels to vertices with similarity measures above a pre-determined threshold value,
\begin{equation}
H(x_i,x_j)\geq h_0.
\label{eq:label_propagation_crit}
\end{equation}
We build a set of benchmark graph instances using software available at \cite{fortunato_site}, which implements the algorithms described in \cite{PhysRevE.80.016118}. The graph order remains fixed at $128$ vertices, and by varying the mean and minimum degree and community size, we generate a set of benchmark graphs with known community memberships. These are randomly connected graphs and the analysis in the remainder of this paper uses only a single instance of each benchmark graph.  In this work we focus on identifying communities in graphs with known ground truths, non-overlapping communities, and each vertex on a graph is only assigned a single community membership. 

The Girvan-Newman benchmark graph \cite{PhysRevE.80.016118,girvan2002community} contains $4$ equal order communities each with $32$ vertices.  It is generated using an average vertex degree $\langle d \rangle = d_{max} = 16$ and $Q_{min}=Q_{max}=32$.  The remaining graphs are generated using the target parameters: $\langle d \rangle = D$ and $Q_{min} = Q_{max} = 2D$. However, depending on the value of $D$ some graph instances $\g{G}_i$ have a larger average degree. The community distributions are summarized in Table \ref{tab:benchmark_graphs}.  
 \begin{table}
 \centering
 \caption{\label{tab:benchmark_graphs} Community distribution and mean degree for benchmark graphs generated on $128$ vertices.}
 \begin{tabular}{l l l l l}
Graph & $|E(\mathcal{G})|$ & $\lbrace (|Q_i|,n_Q)\rbrace$ & $\langle d \rangle $ & D\\
\hline
$\mathcal{G}_A$ & $256$ & $\lbrace(8,16)\rbrace$ & $4$ & $4$\\
$\mathcal{G}_B$ & $512$ & $\lbrace(16,8)\rbrace$ & $8$ & $8$\\
$\mathcal{G}_C$ & $768$ & $\lbrace (24,4),(32,1\rbrace)$ & $12$ & $12$\\
$\mathcal{G}_D$ & $1024$ & $\lbrace(32,4)\rbrace$ & $16$ & $16$\\
$\mathcal{G}_E$ & $1344$ & $\lbrace(42,2), (44,1)\rbrace$ & $21$ & $21$\\
$\mathcal{G}_F$ & $2047$ & $\lbrace(64,2)\rbrace$ & $31.98$ & $32$\\
$\mathcal{G}_G$ & $2959$ & $\lbrace(128,1)\rbrace$ & $46.23$ & $48$\\
$\mathcal{G}_H$ & $3900$ & $\lbrace(128,1)\rbrace$ & $60.93$ & $64$\\
$K_{128}$ & $8128$ & $\lbrace(128,1)\rbrace$ & $127$ & $127$
 \end{tabular}
 \end{table}
The graphs are converted to fully connected spiking neural systems and the resulting dynamics and spiking behavior is simulated using the Python library Brian2 \cite{goodman2008brian} with the following neuron parameters: $v_{th}=0.8 \: \mathrm{V}$, $v_0=v_{R}=0.0 \: \mathrm{V}$, $\tau = 25 \: \mathrm{ms}$, $t_{R}=20 \: \mathrm{ms}$, and synaptic weight amplitude $|s_w|=0.75 \: \mathrm{V}$. 

The square pulse sequence consists of $128$ homogeneous square pulses with shape parameters: $\beta=50.0 \mathrm{s}^{-1}$, $A_{MAX}=12.0 \: \mathrm{V}$, $t_{pulse} = 200 \: \mathrm{ms}$ and subsequent pulses were separated by a gap of $200 \: \mathrm{ms}$. During square pulse driving a neuron would fire a total of $10$ spikes with frequency $\delta_1 \approx 0.21 \: \mathrm{ms}$.  In response, its nearest neighbors will fire a total of $10/2=5$ spikes with frequency $\delta_2 \approx 0.42 \: \mathrm{ms}$. The use of low-order numerical integration methods and approximating a square pulse using a continuous function introduced slight variations in the firing rate, however the total number of spikes fired was not affected. Analogous to randomizing the vertex set in the algorithm of Ref. \cite{raghavan2007near} we randomize the order in which neurons are driven. 

In Ref. \cite{Hamilton_NCAMA} it was demonstrated that the Hamming metric given in Eq. \eqref{eq:Hamming_metric} is linearly separable and can be used to distinguish individual communities. The linear separability of the Hamming metric is dependent on: the fraction of neurons in a system which are driven, and the width $\Delta t$ used for binary decoding.  Our analysis and decoding uses a fixed time step of $\Delta t  = 0.03$ seconds, which is larger than $\delta_1$ but smaller than $\delta_2$.  This time step maximizes the similarity between primary and secondary spiking responses. It is large enough to contain two spikes fired by a neuron being driven by a square pulse, but only one spike fired in response to arriving pulses. 

We construct a spin-glass SNS and drive all neurons with identical square pulses to generate $128$ spike trains. Each neuron ($n_i$)is initialized with a unique label ($y_i^{(0)}=i$), the Hamming similarity threshold ($h_0$) is fixed, and then we update the label of each neuron.  In the algorithm described in \cite{raghavan2007near}, the label of a vertex is updated such that it agrees with the majority of the neighboring vertex labels. Our update rule chooses a neuron to be the source ($x_0$), and its label is fixed ($y_0$). By comparing all neurons to the source, we assign the label $y_0$ to any neuron with similarity measure that satisfies $H(x_0,x_j)\geq h_0$. This is repeated for all $128$ neurons, with each subsequent source neuron chosen at random.  In these initial tests of SLP we only use fully driven neuron sets. 

\begin{figure}[htbp] 
\includegraphics[width=\columnwidth]{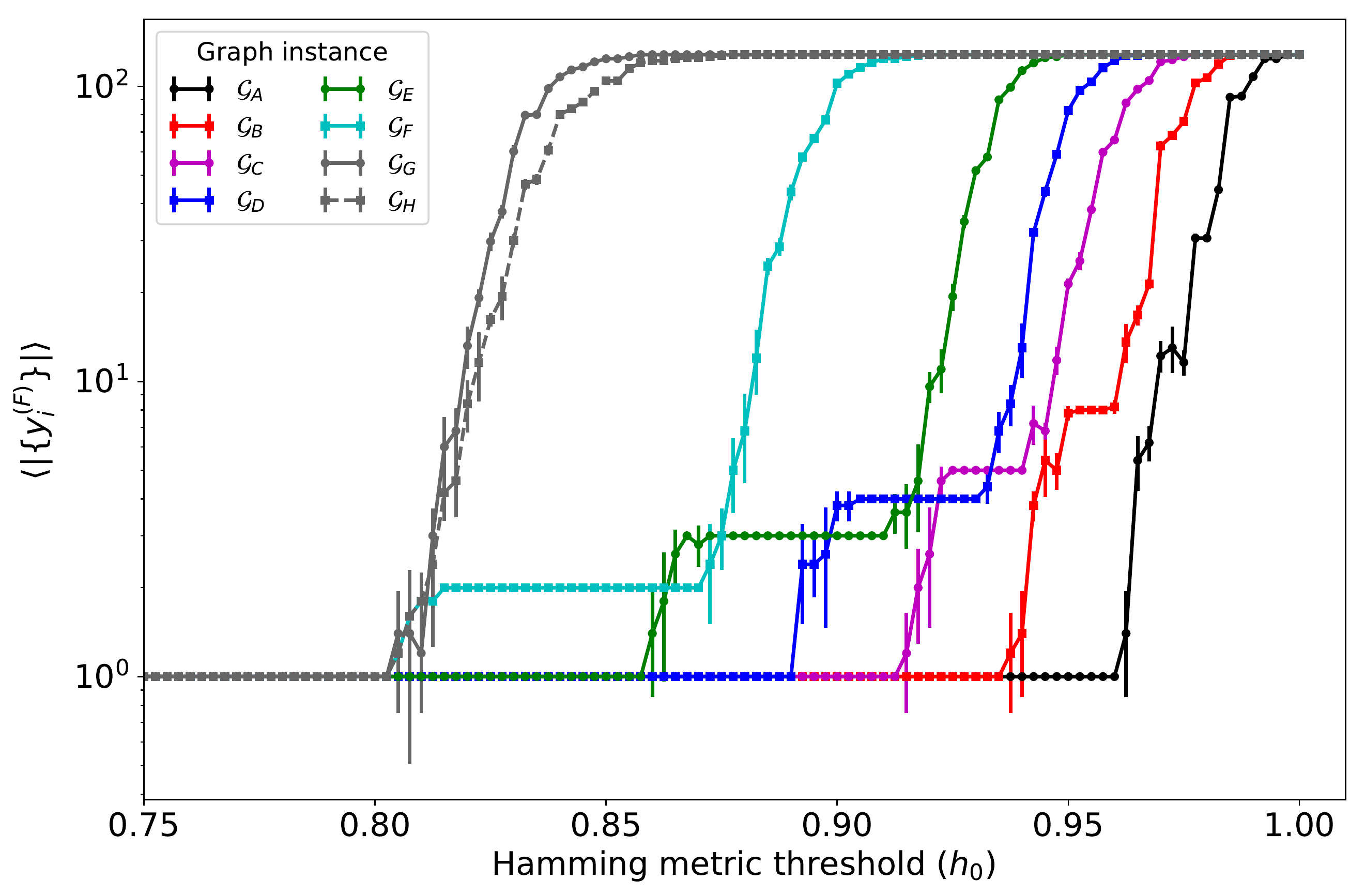}
\caption{(Color online) The mean final size of unique labels, averaged over $5$ iterations of SLP.  For graph instances: $\mathcal{G}_B$ (red), $\mathcal{G}_C$ (magenta), $\mathcal{G}_{D}$ (blue), $\mathcal{G}_E$ (green), and $\mathcal{G}_F$ (cyan), a quasi-stable solution exists between the two trivial solutions. The error bars are defined by the standard deviation.}
\label{fig:label_convergence_Tc}
\end{figure} 
Each graph instance in Table \ref{tab:benchmark_graphs} is generated with a set of ground truth community memberships $A = \lbrace Q_i \rbrace$.  SLP is initialized with $128$ unique labels and after testing each neuron's similarity against every other neuron in the system, the labels will converge to a final label set which is a distribution of community memberships $B = \lbrace Q_j \rbrace$.

In Fig. \ref{fig:label_convergence_Tc} the number of unique labels as a function of $h_0$ is shown for the graphs in Table \ref{tab:benchmark_graphs}. If the Hamming threshold is chosen too low the label set quickly collapses to the trivial solution $\lbrace y_i^{(F)} \rbrace = 1$ and all vertices  are assigned the same label.  If the threshold is chosen too high the label set converges to the trivial solution of $\lbrace y_i^{(F)} \rbrace= [0,127]$; no labels change and each vertex is considered a unique community. The only exception was for a complete graph $K_{128}$, which never deviated from returning a single community label. 

Several graph instances exhibited a quasi-stable solution between the trivial solutions at $|\lbrace y^{(F)}\rbrace|=|\lbrace Q_i\rbrace|$ corresponding to the known number of communities, except for the sparsest graph instance $\mathcal{G}_A$.  For the graph instances $\g{G}_B,\g{G}_C,\g{G}_{D},\g{G}_E,\g{G}_F$, such solutions exist where SLP converges to a non-trivial label set. The range of $h_0$ values that return these solutions indicate that the label assignments are somewhat robust with respect to small deviations in $h_0$. To quantify how well these quasi-stable solutions compare to the known community distributions we use the variance of information $\mathrm{VI}(A,B)$ \cite{meilua2007comparing,ronhovde2010local}, to compare how many vertices in a community $q_k \in \lbrace Q_j \rbrace$ are found in a community $q_l \in \lbrace Q_i \rbrace$, 
\begin{equation}
\mathrm{VI}(A,B) = H(A) + H(B) - 2 MI(A,B).
\end{equation}
Where $H(A), H(B)$ is the Shannon entropy of each distribution, and $MI(A,B)$ measures the mutual information between community memberships $A$ and $B$.  When $\lbrace Q_j \rbrace=\lbrace Q_i \rbrace$ then $\mathrm{VI}(A,B) =0$.  The trivial solutions can also be identified from the variance of information, for the graph instances in Table \ref{tab:benchmark_graphs}, the values of the variance of information is given in Table \ref{tab:VI_benchmark_trivial} for the trivial solutions. For the graph instances with non-trivial solutions, we give the range of $h_0$ where $
\mathrm{VI}(A,B)=0$ in Table \ref{tab:VI_benchmark_ground}.  Graphs $\g{G}_G,\g{G}_H,\g{K}_{128}$ contain only community, reflected by $\mathrm{VI}(A,B)_{1}=0$. 

 \begin{table}
 \centering
 \caption{\label{tab:VI_benchmark_trivial} The variance of information when SLP returns a trivial solution: one single community, or $128$ unique communities.}
 \begin{tabular}{l c c}
Graph & $\mathrm{VI}(A,B)_{1}$ & $\mathrm{VI}(A,B)_{128}$\\
\hline
$\mathcal{G}_A$ & $4.000$ & $3.000$ \\
$\mathcal{G}_B$ & $3.000$ & $4.000$\\
$\mathcal{G}_C$ & $2.311$ & $4.688$\\
$\mathcal{G}_D$ & $2.000$ & $5.000$ \\
$\mathcal{G}_E$ & $1.585$ & $5.415$\\
$\mathcal{G}_F$ & $1.000$ & $6.00$\\
\hline\hline
$\mathcal{G}_G$ & $0.000$ & $7.000$\\
$\mathcal{G}_H$ & $0.000$ & $7.000$\\
$K_{128}$ & $0.000$ & $7.000$
 \end{tabular}
 \end{table}
 
 \begin{table}
 \centering
 \caption{\label{tab:VI_benchmark_ground} For a single instance of SLP, the minimum and maximum values of $h_0$ for which $\mathrm{VI}(A,B)$ vanishes.}
 \begin{tabular}{l c c}
Graph & $\min(h_0)$ & $\max(h_0)$\\
\hline
$\mathcal{G}_B$ & $0.9525$ & $0.9575$\\
$\mathcal{G}_C$ & $0.9250$ & $0.9400$\\
$\mathcal{G}_D$ & $0.9000$ & $0.9325$ \\
$\mathcal{G}_E$ & $0.8650$ & $0.9075$\\
$\mathcal{G}_F$ & $0.8100$ & $0.8725$\\
 \end{tabular}
 \end{table}
 
\begin{figure}[htbp] 
\centering
\includegraphics[width=\columnwidth]{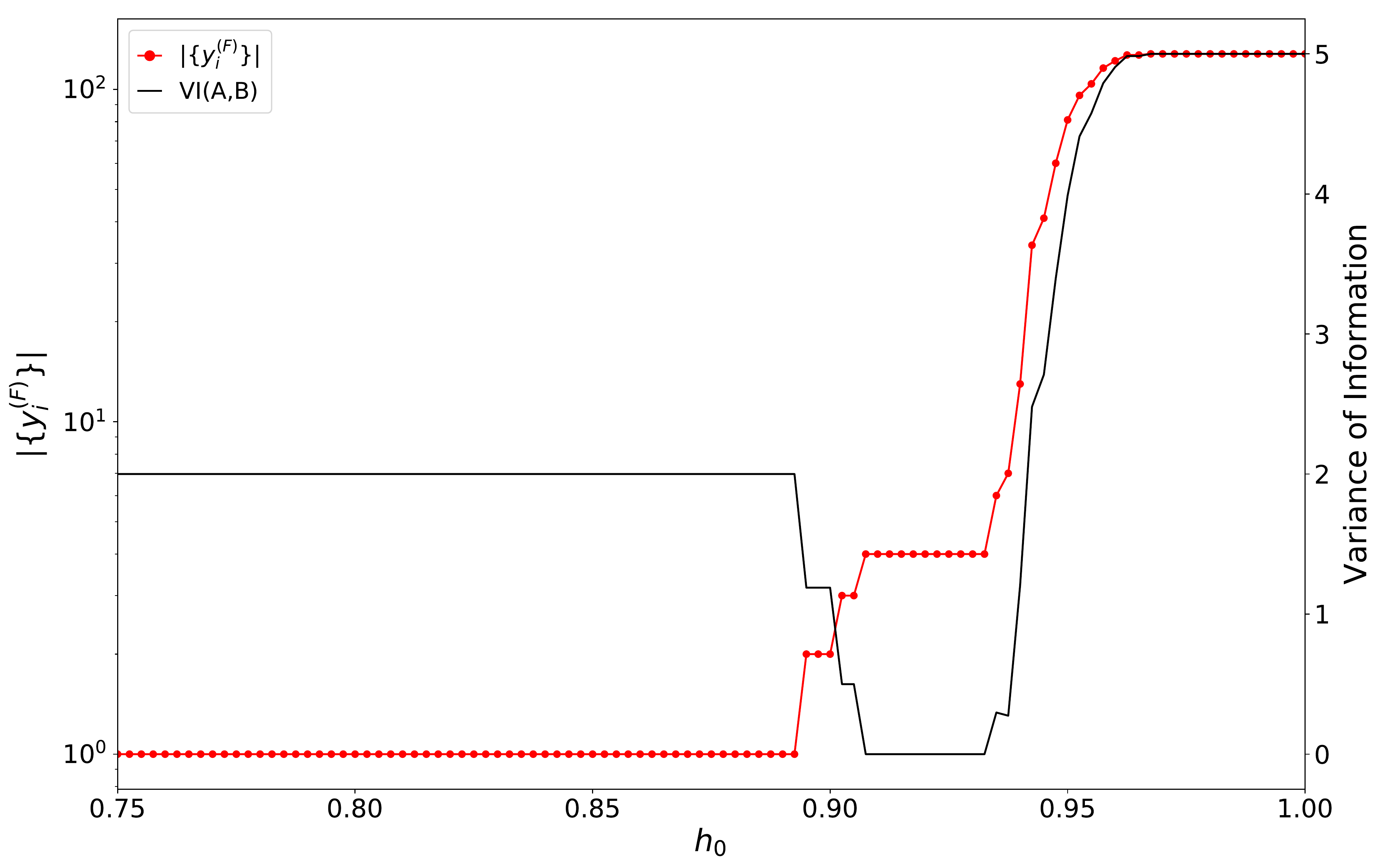}
\caption{For $\mathcal{G}_D$, $VI(A,B) =0$ for the quasi-stable solutions $|\lbrace y_F\rbrace | =4$. Only a single iteration of label propagation is shown.}
\label{fig:label_stability_pulse_width}
\end{figure} 

The convergence to a non-trivial label set, is dependent on the Hamming metric threshold $h_0$ but $h_0$ cannot be used to generate an arbitrary partitioning of $\mathcal{G}$ into $Q$ communities and a value of $h_0$ that returns $Q$ labels many not correspond to an optimal $Q$-partitioning of a graph.  The quasi-stable solutions assign the same label to all vertices in the same community, and distinguishes individual communities by unique labels. In the Appendix \ref{appendix} we show partitions for graph instance $\mathcal{G}_C$ as $h_0$ is swept through a range of $h_0$ values near $|\lbrace y_i^{(F)}\rbrace | = 5$. 

\section{Discussion}
\label{sec:discussion}
The convergence of SLP to non-trivial solutions is dependent on: the SNS mapping and the density of added negative weighted synapses, and the Hamming similarity of Eq. \ref{eq:Hamming_metric}. Section \ref{sec:spike_propagation}, discusses how the fully connected spin glass mapping is too restrictive for graphs with very sparse communities. Section \ref{sec:fine_tuning}, discusses how the choice of embedding and the choice of driving parameters ($\Delta t, h_0, t_A, \delta$) affect the label convergence. In this section we restrict our analysis to what can affect a single iteration of SLP.

\subsection{Graph connectivity and spike propagation}
\label{sec:spike_propagation}
To define a mapping between a graph and a spiking neuron system, we need to identify characteristics that will be useful in identifying strongly connected graph vertices while avoiding loss of information which is encoded in the network topology \cite{scarselli2009graph}.  This section will focus on how local connectivity will determine which neurons fire in response to a driven neuron and in particular we discuss how the fully connected spin glass embedding is overly restrictive for graphs with sparse communities.  Though every edge of the original graph is included in the spiking neuron system, the density of inhibitory synapses causes loss of information about connections in sparse graphs.

 Modularity-based community detection algorithms have a well-known resolution limit, which some spin glass approaches to community detection can avoid \cite{fortunato2007resolution,ronhovde2010local}.  Modularity-based methods tend to fail when trying to find very small communities in very large networks  ($|Q_i| \sim \sqrt{|E(\mathcal{G})|}$).  In Table \ref{tab:benchmark_graphs} $\mathcal{G}_A,\mathcal{G}_B,\mathcal{G}_C$ all have community sizes that fall below this resolution limit.  Yet, only $\mathcal{G}_A$ (the graph with the smallest communities) failed to converge to quasi-stable solutions at $|y^{F}|=|\lbrace Q_i\rbrace|$. We believe that for very sparse communities, the use of fully connected SNS imposes an artificial resolution limit to SLP.

To demonstrate this, we ran SLP on two graph instances $\mathcal{G}_A$ and $\mathcal{G}_D$ with embeddings that are sparser than fully connected spin glasses.  Under the spin-glass mapping, between any pair of neurons there is a length-1 path (a synapse) that is either a positively or negatively weighted.  The inhibition by a single negatively weighted synapse can be mitigated if multiple, positive-weighted length-2 paths also exist between a driven neuron and a neuron which is not connected by a single positive synapse.  To explore the effect of negatively weighted synapses in sparse graphs, we introduced an embedding where inhibitory synapses are only added between pairs of neurons with a high number of length-$2$ paths between them: 
\begin{align}
\label{eq:threshold_embedding}
&\mathrm{if} \: (A^2)_{ij}\leq a_0, \\
&e_{ij} \notin E(\g{G}) \to w_{ij} \in W(\g{S}), s_w < 0.
\end{align}

The minimum variance of information for $\mathcal{G}_A$ is given in Table \ref{tab:alternate_embedding_GA}, and for $\mathcal{G}_D$ is given in Table \ref{tab:alternate_embedding_GD}.  We tested several threshold values and also the direct mapping of a graph in which zero inhibitory synapses are used.  The sparser graph $\mathcal{G}_A$ showed an improvement in community identification for sparser embeddings, which the denser graph $\mathcal{G}_D$ continued to return the exact ground truth.  Neither graph was able to return the ground truth memberships when using the direct mapping, in fact $\mathcal{G}_D$ returned the trivial solution of only $1$ community.

\begin{table}
 \centering
 \caption{\label{tab:alternate_embedding_GA}The variance of information alternate embeddings with $\mathcal{G}_A$.}
 \begin{tabular}{l c c}
Embedding & $|s_w|<0$ & $\min[VI(A,B)]$ \\
\hline
Fully connected & 7872 & $0.583$ \\
$a_0 = 3$ & $760$ & $0.399$\\
$a_0 = 2$ & $694$ & $0.307$\\
$a_0 = 1$ & $456$ & $0.000$ \\
Direct Map & $0$ & $0.0679$
 \end{tabular}
 \end{table}
 
\begin{table}
 \centering
 \caption{\label{tab:alternate_embedding_GD}The variance of information alternate embeddings with $\mathcal{G}_D$.}
 \begin{tabular}{l c c}
Embedding & $|s_w|<0$ & $\min[VI(A,B)]$ \\
\hline
Fully connected & 7104 & $0.00$ \\
$a_0 = 3$ & $5126$ & $0.00$\\
$a_0 = 2$ & $5026$ & $0.00$\\
$a_0 = 1$ & $4256$ & $0.00$ \\
Direct Map & $0$ & $2.0$\\
\end{tabular}
 \end{table}

\subsection{Neuron parameters and Hamming similarity}
\label{sec:fine_tuning}
The performance of SLP depends on the similarity of the binary decoded spike trains, which is quantified by the Hamming distance in Eq. \ref{eq:Hamming_metric}. Additionally, the choice of Hamming metric threshold ($h_0$) controls the ability to distinguish each community. The linear separability of Eq. \eqref{eq:Hamming_metric} was previously discussed in Section \ref{sec:spiking_label_propagation} with respect to the size of the decoding window $\Delta t$.  It was found that choosing a time window of $\Delta t$ that satisfies $\delta_1 < \Delta t < 2\delta_1 \ll t_A$ will minimize the distance between the binary vector of a driven neuron and any neuron spiking in response.  This fixes the length of each vector at $L = T/\Delta t$, where $T=t_f-t_0$ is defined by: $t_f$, the time the final neuron fires its final spike and $t_0$ the time that the first neuron fires its first spike.  

It is observed in Fig. \ref{fig:label_convergence_Tc} that as the number of communities in the graph grows, and the relative size of each community becomes smaller in relation to the overall graph order, the non-trivial solutions become less stable.  It can be inferred that the sparsity of the graph will affect the spiking behavior by reducing the number of neurons that fire in response to a driven neuron. For a short pulse, only a few spikes will be fired by a driven neuron, and fewer will be fired by neighboring neurons. These sparse spike trains will result in binary vectors $x_i$ with very few nonzero entries and the term $h(x_i,x_j)/L$ can become very small.

\begin{figure}[htbp] 
\centering
\includegraphics[width=\columnwidth]{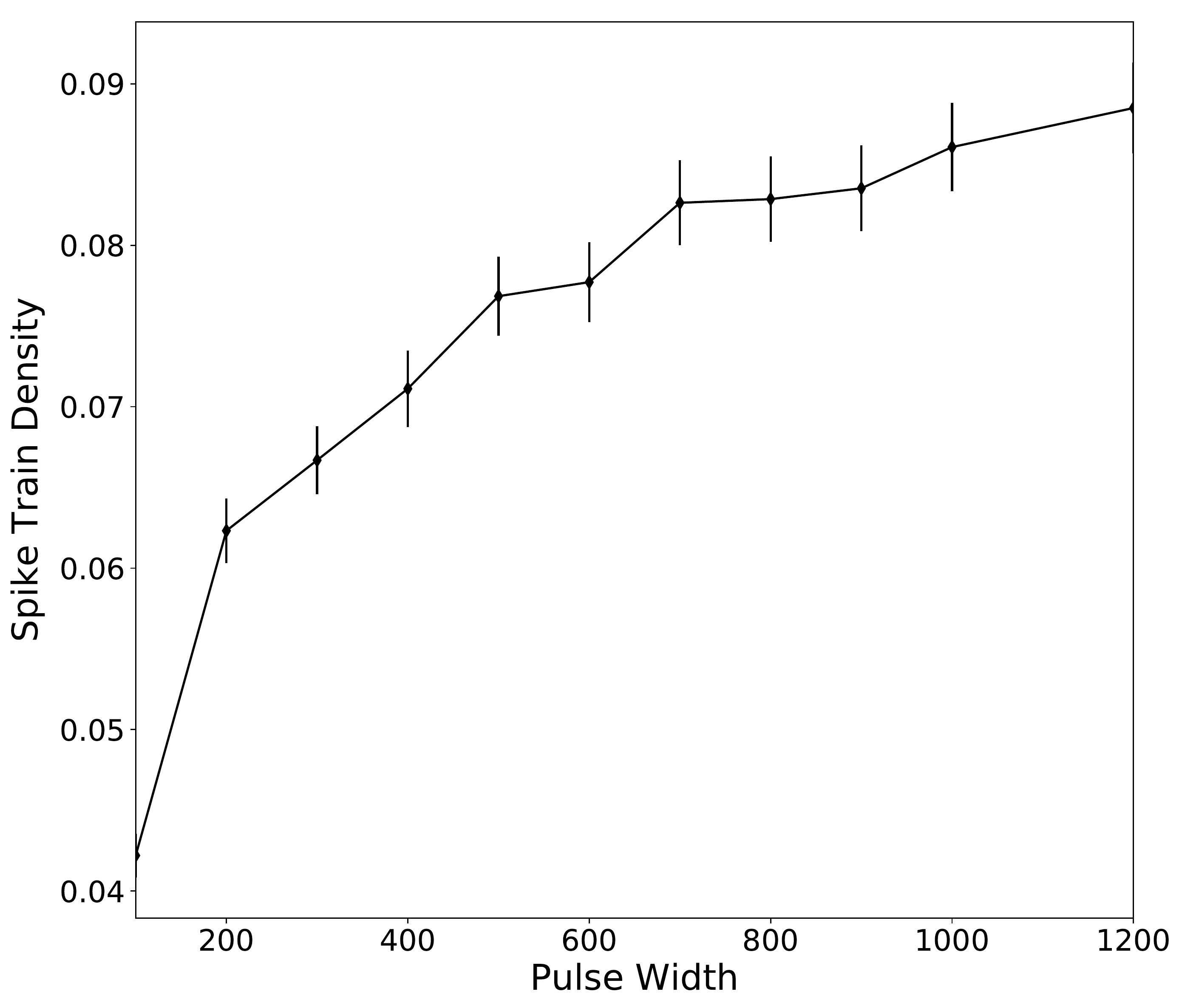}
\caption{For benchmark graph $\mathcal{G}_D$ the weight $\langle |x_i|_1/L \rangle$ averaged over all neurons, increases as the pulse width increases, while the gap between pulses remains constant ($\delta = 200 \: \mathrm{ms}$). Each spike train was decoded using a time window width $\Delta t = 0.03 \: \mathrm{s}$. The standard deviation defines the error bars.}
\label{fig:varwidth_density}
\end{figure} 

\begin{figure}[htbp] 
\centering
\includegraphics[width=\columnwidth]{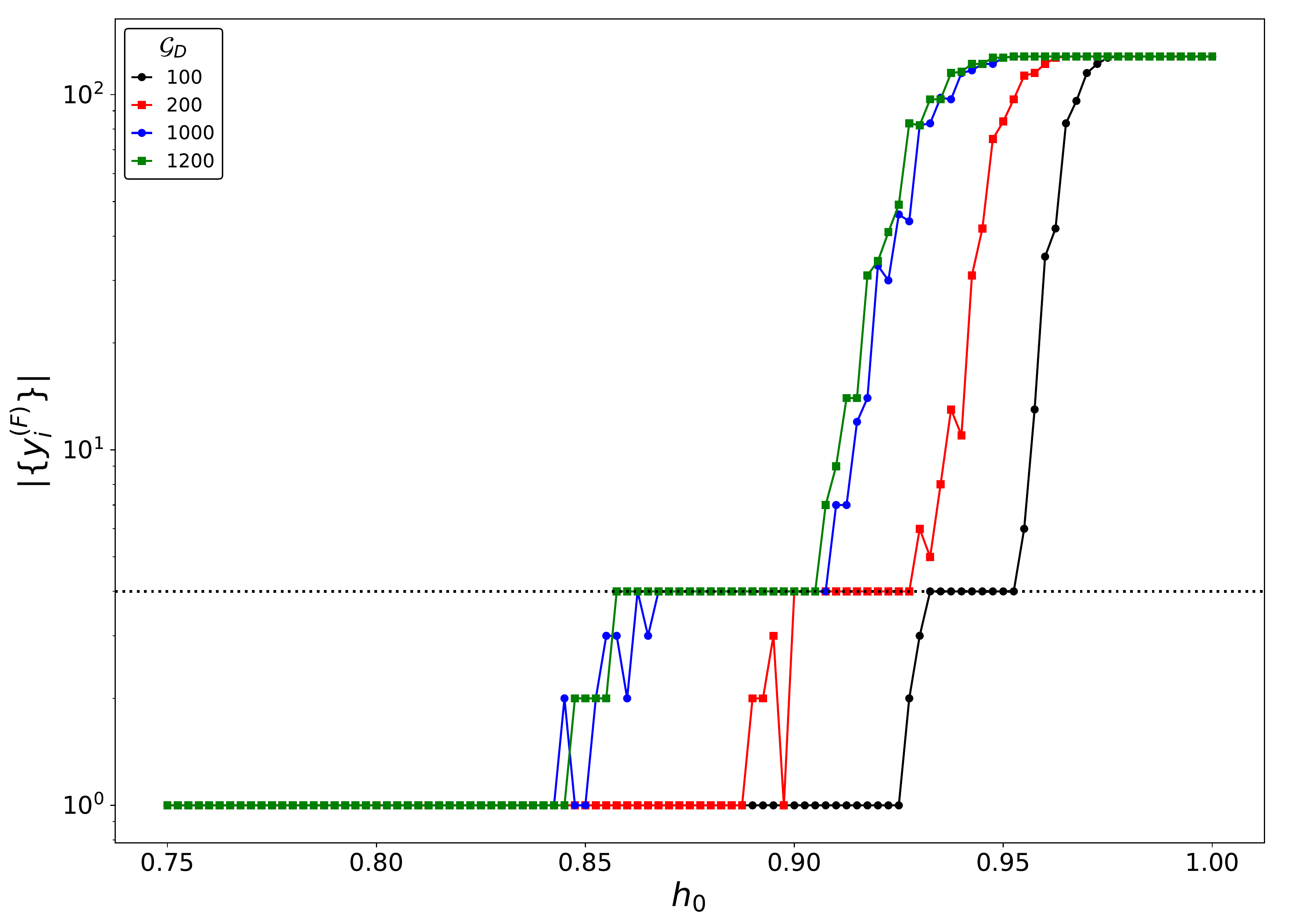}
\caption{(Color online) For a single instance of label propagation on graph instance $\mathcal{G}_D$: the final size of unique labels after label propagation terminates for pulse widths $t_A=[100,200,1000,1200] \: \mathrm{ms}$ while the gap between pulses remains constant ($\delta = 200 \: \mathrm{ms}$).}
\label{fig:label_convergence_pulse_width}
\end{figure} 
To compensate for this spike reduction, we can increase the length of the driving pulse $t_A$ which will change the Hamming distance between two vectors ($x_i,x_j$) by increasing or decreasing $|x_i|_1$, while holding $|x_j|_1$ constant (or vice versa).  In Fig. \ref{fig:varwidth_density}, increasing the length of $t_A$ leads to denser spike trains.  Its effect on the overall label set convergence is that a longer pulse can lead to more stable label sets (see Fig. \ref{fig:label_convergence_pulse_width}).  However, this will also increase the overall duration of the spiking dynamics simulation, and in turn increase the vector length which can reduce the efficacy of the term $h(x_i,x_j)/L$ in Eq. \eqref{eq:Hamming_metric}.  

To minimize the increase in the vector length $L$, and reduce the number of $0$ entries in all vectors, then $\delta$ must be kept short, but cannot be arbitrarily set.  We do not incorporate a floor on the potential value, instead we require the time constant $\tau$ to be small enough that any effects incurred during a square pulse have effectively decayed away before the next pulse is applied: each neuron in a system is always assumed to be within $10\%$ of the resting voltage $v_0$ when a square pulse is applied to any neuron. There is a limit to how small $\delta$ can be, decreasing the gap between pulses requires that $\tau$ becomes smaller so any voltage difference incurred from arriving spikes must completely decay away.  Yet as $\tau$ decreases, this requires an increase in the synaptic weight in order to ensure the correlated spike dynamics are generated.  

To demonstrate this, we ran multiple SLP instances on $\g{G}_A$ with various values for $\tau, \delta, \alpha$ while keeping $v_{th},A_{MAX}$ and $t_A$ constant.  To ensure the localized spiking behavior is generated we impose two constraints on the parameters:
\begin{equation}
    (s_w)(1+e^{-\delta_1/\tau})\geq v_{th},
    \label{eq:condition1}
\end{equation}
\begin{equation}
    |(2-\langle D \rangle)(e^{-\delta/\tau})|\leq 0.01.
    \label{eq:condition2}
\end{equation}
The condition in Eq. \ref{eq:condition1} ensures that $2$ spikes are sufficient to cause a neuron to spike and is found by integrating the equation of motion in Eq. \ref{eq:LIF_EOM}.  The second condition is a heuristic that ensures that any change in a neuron's potential has sufficiently decayed away before a new pulse is applied to a system.  We assume a lower bound on the suppressed neuron potential of $(2-\langle D \rangle)$, from a neuron which is completely disconnected from a driven neuron and its nearest neighbors.  For the sparsest graph instance $\g{G}_A$ this was sufficient, however we do not expect this lower bound to hold for more densely connected graphs.
\begin{figure}[htbp] 
\centering
\includegraphics[width=0.95\columnwidth]{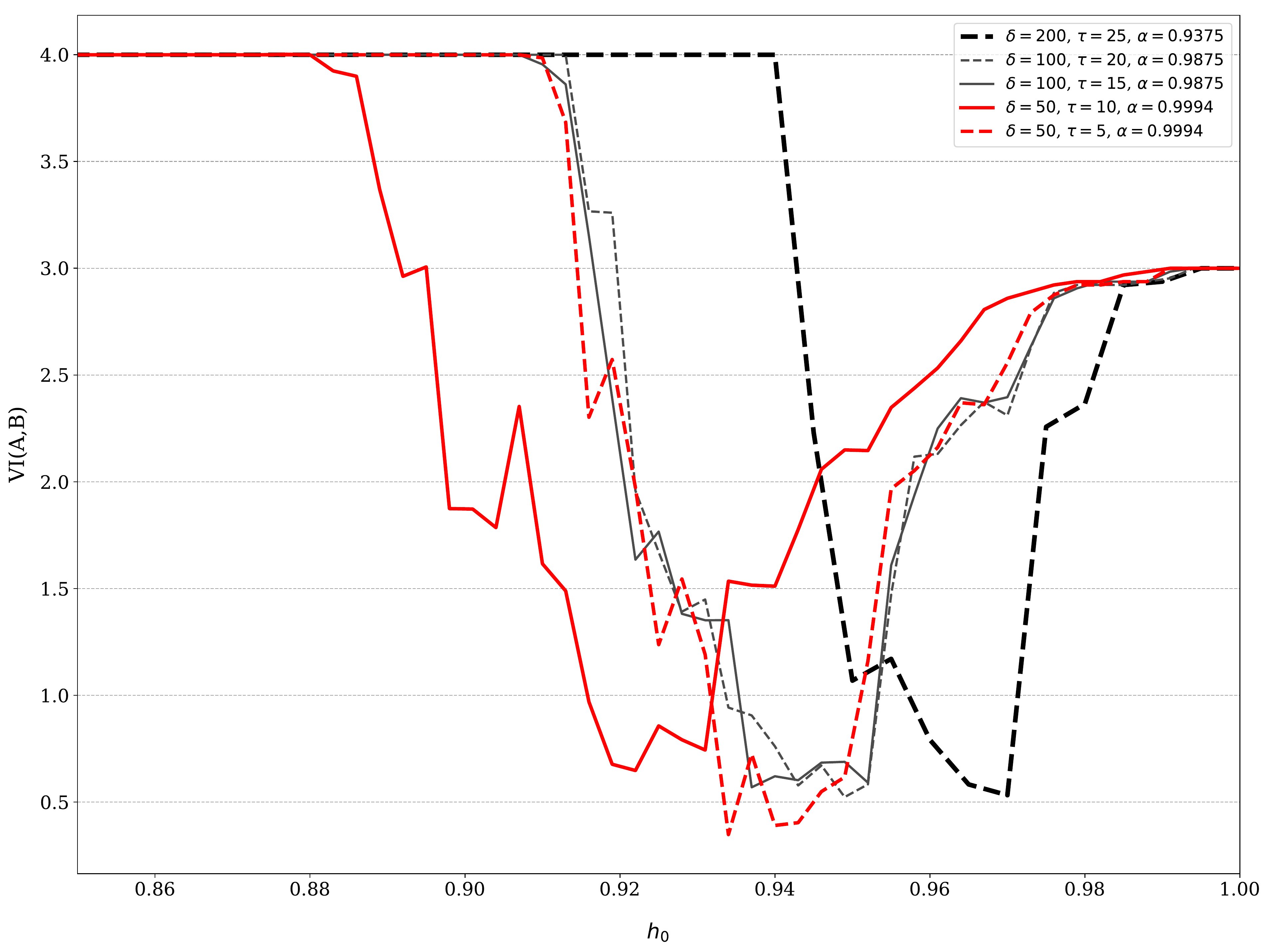}
\caption{(Color online) For a single instance of label propagation on graph instance $\mathcal{G}_A$: the variance of information after label propagation terminates for gap widths $\delta=[200,100,50] \: \mathrm{ms}$ while the pulse width remains constant ($t_A = 200 \: \mathrm{ms}$).  As $\delta$ is reduced, the parameters $\tau,\alpha$ are scaled accordingly to ensure localized spiking behavior. For all plotted parameters, $\min{(VI)}=0.3474$ for $\delta = 50, \tau = 5, \alpha = 0.9994$.}
\label{fig:tau_var_plot}
\end{figure} 

\section{Conclusions}
\label{sec:conclusions}
In this work we have described SLP, a label propagation method that incorporates spiking neurons by mapping a graph $\mathcal{G}(V,E)$ to a fully-connected SNS $\mathcal{S}(N,W)$. The convergence of SLP to the ground truth community memberships is dependent on the similarity between neuron spiking behavior This current implementation of SLP relies on static synapses and label propagation is implemented after all neurons in $\g{S}$ are driven.  In Section \ref{sec:spiking_label_propagation}, we showed that a fully connected SNS was able to return the ground truth  memberships for graphs with dense communities, however this mapping was overly restrictive for graphs with sparse communities.  

Our goal in deriving SLP was to find an algorithm for spiking neurons that did not require extensive pre-training in order to set the parameters of $\mathcal{S}$.  In SLP, a graph is used to define the connections of a system of spiking neurons with parameters that can be determined from closed form analytic expressions \cite{Hamilton_NCAMA}.  The spiking neurons are used to quickly generate spiking dynamics used in later analysis.  In this work, the generation of spiking data is done through numerical simulation of nonlinear neuron dynamics, but SLP has been adapted and deployed on neuromorphic hardware \cite{Hamilton_JETC}.  

Conventional label propagation updates the label for a given vertex by comparing to the labels of the nearest neighboring vertices. Our spiking implementation of label propagation updates the labels for a set of vertices depending on the similarity of their associated spike trains with the spike train associated with a given vertex.  Conventional label propagation may require several iterations through the vertex set before it converges to a final partition.  Our spiking label propagation can converge to a stable set of labels within a few label updates, or it can require iterating through the entire neuron set once.

Future development of SLP is focused on incorporating synaptic plasticity.  This will allow for the weights of $\mathcal{S}$ to change.  This would incorporate more spiking neuron dynamics, and can improve performance on graphs with communities of varying sparsity. Recently, an initial study in to developing adaptive SLP methods has been implemented. \cite{Hamilton_ICONS}.

\section*{acknowledgments}
This work was supported in part by the United States Department of Defense and used resources of the Computational Research and Development Programs at Oak Ridge National Laboratory.  Research sponsored in part by the Laboratory Directed Research and Development Program of Oak Ridge National Laboratory, managed by UT-Battelle, LLC, for the U. S. Department of Energy.

\bibliography{bib_files/PRE_lit,bib_files/neuromorph_2018,bib_files/graph_alg_lit,bib_files/neuromorphic_lit}

\begin{thebibliography}{10}

\bibitem{maass1997networks}
Wolfgang Maass.
\newblock Networks of spiking neurons: the third generation of neural network
  models.
\newblock {\em Neural Networks}, 10(9):1659--1671, 1997.

\bibitem{maass2001pulsed}
Wolfgang Maass and Christopher~M Bishop.
\newblock {\em Pulsed Neural Networks}.
\newblock MIT Press, 2001.

\bibitem{indiveri2015memory}
Giacomo Indiveri and Shih-Chii Liu.
\newblock Memory and information processing in neuromorphic systems.
\newblock {\em Proceedings of the IEEE}, 103(8):1379--1397, 2015.

\bibitem{merolla2014million}
Paul~A Merolla, John~V Arthur, Rodrigo Alvarez-Icaza, Andrew~S Cassidy, Jun
  Sawada, Filipp Akopyan, Bryan~L Jackson, Nabil Imam, Chen Guo, Yutaka
  Nakamura, et~al.
\newblock A million spiking-neuron integrated circuit with a scalable
  communication network and interface.
\newblock {\em Science}, 345(6197):668--673, 2014.

\bibitem{painkras2013spinnaker}
Eustace Painkras, Luis~A Plana, Jim Garside, Steve Temple, Francesco Galluppi,
  Cameron Patterson, David~R Lester, Andrew~D Brown, and Steve~B Furber.
\newblock Spinnaker: A 1-w 18-core system-on-chip for massively-parallel neural
  network simulation.
\newblock {\em IEEE Journal of Solid-State Circuits}, 48(8):1943--1953, 2013.

\bibitem{davies2018loihi}
Mike Davies, Narayan Srinivasa, Tsung-Han Lin, Gautham Chinya, Yongqiang Cao,
  Sri~Harsha Choday, Georgios Dimou, Prasad Joshi, Nabil Imam, Shweta Jain,
  et~al.
\newblock Loihi: A neuromorphic manycore processor with on-chip learning.
\newblock {\em IEEE Micro}, 38(1):82--99, 2018.

\bibitem{esser2015backpropagation}
Steve~K Esser, Rathinakumar Appuswamy, Paul Merolla, John~V Arthur, and
  Dharmendra~S Modha.
\newblock Backpropagation for energy-efficient neuromorphic computing.
\newblock In {\em Advances in Neural Information Processing Systems}, pages
  1117--1125, 2015.

\bibitem{stromatias2015scalable}
Evangelos Stromatias, Daniel Neil, Francesco Galluppi, Michael Pfeiffer,
  Shih-Chii Liu, and Steve Furber.
\newblock Scalable energy-efficient, low-latency implementations of trained
  spiking deep belief networks on spinnaker.
\newblock In {\em Neural Networks (IJCNN), 2015 International Joint Conference
  on}, pages 1--8. IEEE, 2015.

\bibitem{das2015gibbs}
Srinjoy Das, Bruno~Umbria Pedroni, Paul Merolla, John Arthur, Andrew~S Cassidy,
  Bryan~L Jackson, Dharmendra Modha, Gert Cauwenberghs, and Ken Kreutz-Delgado.
\newblock Gibbs sampling with low-power spiking digital neurons.
\newblock In {\em 2015 IEEE International Symposium on Circuits and Systems
  (ISCAS)}, pages 2704--2707. IEEE, 2015.

\bibitem{esser2016convolutional}
Steven~K Esser, Paul~A Merolla, John~V Arthur, Andrew~S Cassidy, Rathinakumar
  Appuswamy, Alexander Andreopoulos, David~J Berg, Jeffrey~L McKinstry, Timothy
  Melano, Davis~R Barch, et~al.
\newblock Convolutional networks for fast, energy-efficient neuromorphic
  computing.
\newblock {\em Proceedings of the National Academy of Sciences}, page
  201604850, 2016.

\bibitem{pedroni2016mapping}
Bruno~U Pedroni, Srinjoy Das, John~V Arthur, Paul~A Merolla, Bryan~L Jackson,
  Dharmendra~S Modha, Kenneth Kreutz-Delgado, and Gert Cauwenberghs.
\newblock Mapping generative models onto a network of digital spiking neurons.
\newblock {\em IEEE Transactions on Biomedical Circuits and Systems},
  10(4):837--854, 2016.

\bibitem{diehl2016conversion}
Peter~U Diehl, Guido Zarrella, Andrew Cassidy, Bruno~U Pedroni, and Emre
  Neftci.
\newblock Conversion of artificial recurrent neural networks to spiking neural
  networks for low-power neuromorphic hardware.
\newblock In {\em IEEE International Conference on Rebooting Computing (ICRC)},
  pages 1--8. IEEE, 2016.

\bibitem{constraintSNN_Maass_2016}
Zeno Jonke, Stefan Habenschuss, and Wolfgang Maass.
\newblock Solving constraint satisfaction problems with networks of spiking
  neurons.
\newblock {\em Frontiers in Neuroscience}, 10:118, 2016.

\bibitem{constraintSNN_SpiNNaker_2017}
Gabriel~Andr{\'e}s Fonseca~Guerra and Steve~B Furber.
\newblock Using stochastic spiking neural networks on spinnaker to solve
  constraint satisfaction problems.
\newblock {\em Frontiers in neuroscience}, 11:714, 2017.

\bibitem{severa2018spiking}
William Severa, Rich Lehoucq, Ojas Parekh, and James~B Aimone.
\newblock Spiking neural algorithms for markov process random walk.
\newblock {\em arXiv preprint arXiv:1805.00509}, 2018.

\bibitem{severa2016spiking}
William Severa, Ojas Parekh, Kristofor~D Carlson, Conrad~D James, and James~B
  Aimone.
\newblock Spiking network algorithms for scientific computing.
\newblock In {\em Rebooting Computing (ICRC), IEEE International Conference
  on}, pages 1--8. IEEE, 2016.

\bibitem{boccaletti2006complex}
Stefano Boccaletti, Vito Latora, Yamir Moreno, Martin Chavez, and D-U Hwang.
\newblock Complex networks: Structure and dynamics.
\newblock {\em Physics Reports}, 424(4):175--308, 2006.

\bibitem{fortunato2010community}
Santo Fortunato.
\newblock Community detection in graphs.
\newblock {\em Physics Reports}, 486(3):75--174, 2010.

\bibitem{malliaros2013clustering}
Fragkiskos~D Malliaros and Michalis Vazirgiannis.
\newblock Clustering and community detection in directed networks: A survey.
\newblock {\em Physics Reports}, 533(4):95--142, 2013.

\bibitem{schaub2017many}
Michael~T Schaub, Jean-Charles Delvenne, Martin Rosvall, and Renaud Lambiotte.
\newblock The many facets of community detection in complex networks.
\newblock {\em Applied Network Science}, 2(1):4, 2017.

\bibitem{xu2005survey}
Rui Xu and Donald Wunsch.
\newblock Survey of clustering algorithms.
\newblock {\em IEEE Transactions on neural networks}, 16(3):645--678, 2005.

\bibitem{raghavan2007near}
Usha~Nandini Raghavan, R{\'e}ka Albert, and Soundar Kumara.
\newblock Near linear time algorithm to detect community structures in
  large-scale networks.
\newblock {\em Physical Review E}, 76(3):036106, 2007.

\bibitem{barber2009detecting}
Michael~J Barber and John~W Clark.
\newblock Detecting network communities by propagating labels under
  constraints.
\newblock {\em Physical Review E}, 80(2):026129, 2009.

\bibitem{tibely2008equivalence}
Gergely Tib{\'e}ly and J{\'a}nos Kert{\'e}sz.
\newblock On the equivalence of the label propagation method of community
  detection and a potts model approach.
\newblock {\em Physica A: Statistical Mechanics and its Applications},
  387(19):4982--4984, 2008.

\bibitem{PhysRevLett.76.3251}
Marcelo Blatt, Shai Wiseman, and Eytan Domany.
\newblock Superparamagnetic clustering of data.
\newblock {\em Phys. Rev. Lett.}, 76:3251--3254, Apr 1996.

\bibitem{PhysRevE.74.016110}
J\"org Reichardt and Stefan Bornholdt.
\newblock Statistical mechanics of community detection.
\newblock {\em Phys. Rev. E}, 74:016110, Jul 2006.

\bibitem{reichardt2004detecting}
J{\"o}rg Reichardt and Stefan Bornholdt.
\newblock Detecting fuzzy community structures in complex networks with a
  {P}otts model.
\newblock {\em Phys. Rev. Lett.}, 93(21):218701, 2004.

\bibitem{quiles2010label}
Marcos~G Quiles, Liang Zhao, Fabricio~A Breve, and Anderson Rocha.
\newblock Label propagation through neuronal synchrony.
\newblock In {\em Neural Networks (IJCNN), The 2010 International Joint
  Conference on}, pages 1--8. IEEE, 2010.

\bibitem{quiles2013dynamical}
Marcos~G Quiles, Ezequiel~R Zorzal, and Elbert~EN Macau.
\newblock A dynamical model for community detection in complex networks.
\newblock In {\em Neural Networks (IJCNN), The 2013 International Joint
  Conference on}, pages 1--8. IEEE, 2013.

\bibitem{de2014community}
Jo{\~a}o Eliakin~Mota de~Oliveira and Marcos~G Quiles.
\newblock Community detection in complex networks using coupled kuramoto
  oscillators.
\newblock In {\em Computational Science and Its Applications (ICCSA), 2014 14th
  International Conference on}, pages 85--90. IEEE, 2014.

\bibitem{Lowel209}
S~Lowel and W~Singer.
\newblock Selection of intrinsic horizontal connections in the visual cortex by
  correlated neuronal activity.
\newblock {\em Science}, 255(5041):209--212, 1992.

\bibitem{cassidy2013cognitive}
Andrew~S Cassidy, Paul Merolla, John~V Arthur, Steve~K Esser, Bryan Jackson,
  Rodrigo Alvarez-Icaza, Pallab Datta, Jun Sawada, Theodore~M Wong, Vitaly
  Feldman, et~al.
\newblock Cognitive computing building block: A versatile and efficient digital
  neuron model for neurosynaptic cores.
\newblock In {\em The 2013 International Joint Conference on Neural Networks
  (IJCNN)}, pages 1--10. IEEE, 2013.

\bibitem{Hamilton_NCAMA}
Kathleen~E. Hamilton, Neena Imam, and Travis~S. Humble.
\newblock Community detection with spiking neural networks for neuromorphic
  hardware.
\newblock In {\em Proceedings of the Neuromorphic Computing Symposium}, NCS
  '17, pages 9:1--9:8, New York, NY, USA, 2017. ACM.

\bibitem{humphries2011spike}
Mark~D Humphries.
\newblock Spike-train communities: finding groups of similar spike trains.
\newblock {\em Journal of Neuroscience}, 31(6):2321--2336, 2011.

\bibitem{fortunato_site}
Santo Fortunato.
\newblock Santo {F}ortunato's website: Software.
\newblock \url{https://sites.google.com/site/santofortunato/inthepress2}, 2017.
\newblock Accessed: 2017-05-10.

\bibitem{PhysRevE.80.016118}
Andrea Lancichinetti and Santo Fortunato.
\newblock Benchmarks for testing community detection algorithms on directed and
  weighted graphs with overlapping communities.
\newblock {\em Phys. Rev. E}, 80:016118, Jul 2009.

\bibitem{girvan2002community}
Michelle Girvan and Mark~EJ Newman.
\newblock Community structure in social and biological networks.
\newblock {\em Proceedings of the National Academy of Sciences},
  99(12):7821--7826, 2002.

\bibitem{goodman2008brian}
Dan Goodman and Romain Brette.
\newblock Brian: A simulator for spiking neural networks in python.
\newblock {\em Frontiers in Neuroinformatics}, 2, 2008.

\bibitem{meilua2007comparing}
Marina Meil{\u{a}}.
\newblock Comparing clusterings-- an information based distance.
\newblock {\em Journal of multivariate analysis}, 98(5):873--895, 2007.

\bibitem{ronhovde2010local}
Peter Ronhovde and Zohar Nussinov.
\newblock Local resolution-limit-free potts model for community detection.
\newblock {\em Physical Review E}, 81(4):046114, 2010.

\bibitem{scarselli2009graph}
Franco Scarselli, Marco Gori, Ah~Chung Tsoi, Markus Hagenbuchner, and Gabriele
  Monfardini.
\newblock The graph neural network model.
\newblock {\em IEEE Transactions on Neural Networks}, 20(1):61--80, 2009.

\bibitem{fortunato2007resolution}
Santo Fortunato and Marc Barthelemy.
\newblock Resolution limit in community detection.
\newblock {\em Proceedings of the National Academy of Sciences}, 104(1):36--41,
  2007.

\bibitem{Hamilton_JETC}
Kathleen~E. Hamilton, Neena Imam, and Travis~S. Humble.
\newblock Sparse hardware embedding of spiking neuron systems for community
  detection.
\newblock {\em JETC: SI- Neuromorphic Computing}, 1:1, 2018.
\newblock to appear.

\bibitem{Hamilton_ICONS}
Kathleen~E. Hamilton and Catherine~D. Schuman.
\newblock Towards adaptive spiking label propagation.
\newblock In {\em Proceedings of the International Conference on Neuromorphic
  Systems}, ICONS '18, pages 13:1--13:8, New York, NY, USA, 2018. ACM.

\end{thebibliography}

\appendices
\section{Sample partitioning of a graph}
\label{appendix}
The quasi-stable solutions in Fig. \ref{fig:label_convergence_Tc} correspond to values of $h_0$ that return a partitioning of the graph instance $\mathcal{G}_i$ into communities with the following characteristics: all vertices in the same community have the same label, and each community has a unique label. While it is possible to return a final label set with an arbitrary number of labels, this may not correspond to an optimal partitioning of the graph, which we define in terms of the known ground truths.We demonstrate this using graph instance $\mathcal{G}_C$, which has $5$ communities.  

\begin{figure}
    \centering
    \includegraphics[width=0.3\textwidth]{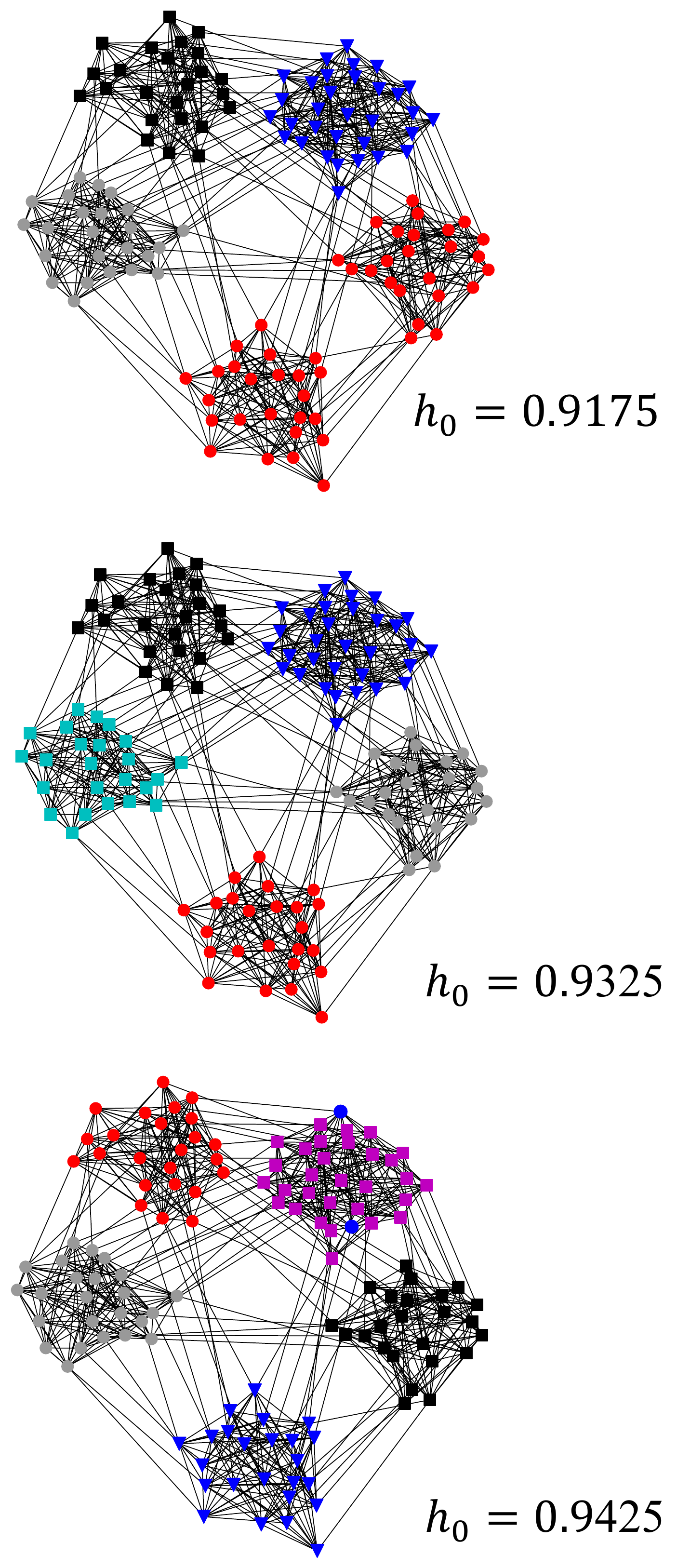}
    \caption{(Color online) Final label sets for $\mathcal{G}_C$ for various threshold values.}\label{fig:GC_partitioning}
\end{figure}

When $h_0=0.9175$, SLP terminates with only $4$ unique labels. The partitioning of $\mathcal{G}_C$ with only $4$ assigns the same label to all vertices in the same community, but does not uniquely distinguish all communities (see Fig. \ref{fig:GC_partitioning}(a)).
When $h_0=0.9325$, the label propagation algorithm terminates with $5$ unique labels. The partitioning of $\mathcal{G}_C$ with $5$ labels uniquely distinguishes all communities and uniformly identifies all vertices in the same community (see Fig. \ref{fig:GC_partitioning}(b)). Each label in the final label set overlaps exactly with a label in the known label set that overlaps.
When $h_0=0.9425$, the label propagation algorithm terminates with $6$ unique labels. The partitioning of $\mathcal{G}_C$ with $6$ labels does not return a $6$-partitioning, rather it returns the $5$ communities from the $h_0=0.9325$ partitioning, with several mislabeled vertices (see Fig. \ref{fig:GC_partitioning}(c)).

\end{document}